# Telluride nanocrystals with adjustable amorphous shell thickness and core-shell structure modulation by aqueous cation-exchange

Xinyuan Li[a, b, +], Mengyao Su[a, +], Yi-Chi Wang[c, d +], Meng Xu[a, *], Minman Tong[e, *], Sarah J. Haigh[c], Jiatao Zhang[a, b, *]

a.  Beijing Key Laboratory of Construction-Tailorable Advanced Functional Materials and Green Applications, School of Materials, Beijing Institute of Technology, Beijing 100081, China. E-mail: zhangjt@bit.edu.cn; xumeng@bit.edu.cn

b.  School of Chemistry, Beijing Institute of Technology, Beijing 100081.

c.  Department of Materials, University of Manchester, Manchester M13 9PL, U.K.

d.  Beijing Institute of Nanoenergy and Nanosystems, Chinese Academy of Sciences, Beijing 101400, China

e.  School of chemistry and materials science, Jiangsu Normal University, Xuzhou 221116, China. Email: tongmm@jsnu.edu.cn

+   These authors contributed equally to this work.

## Abstract:

Engineering the structure of core-shell colloidal semiconductor nanoparticles (CSNPs) is attractive due to the potential to enhance photo-induced charge transfer (PICT) and induce favourable optical and electronic properties. Nonetheless, the sensitivity of telluride CSNPs to high temperatures makes it challenging to precisely modulate their surface crystallinity. Herein, we have developed an efficient strategy for synthesising telluride CSNPs with thin amorphous shells using aqueous cation exchange (ACE). By changing the synthesis temperature in the range 40 to 110 °C, the crystallinity of the CdTe nanoparticles was controllable from perfect crystals with no detectable amorphous shell (c-CdTe) to a core-shell structure with a crystalline CdTe NP core covered by an amorphous shell of tunable thickness up to 7-8nm (c@a-CdTe) . A second ACE step transformed the c@a-CdTe to crystalline CdTe@HgTe core-shell NPs. The

c@a-CdTe nanoparticles synthesized at 60°C and having a 4-5 nm thick amorphous shell, exhibited the highest surface-enhanced Raman scattering activity with a high enhancement factor around 8.82×10$^5$, attributed to the coupling between the amorphous shell and the crystalline core.

**Introduction**

The crystallinity and surface/interface atomic structure of colloidal semiconductor nanocrystals (CSNPs) play important roles in determining their optical, optoelectronic and catalytic properties[1]. In the last decades, many strategies have been reported for the synthesis of CSNPs, among which cation exchange reactions (CERs) have been identified as an efficient and facile strategy for the precise synthesis of complex semiconductor nanoparticles (NPs) with desirable properties[2]. The thermodynamics and kinetics of CER can be modified by adjusting surface ligands, solvents, cations and reaction temperatures, etc., which provides a large parameter space for tailoring the composition and surface condition of the as-produced CSNPs[3]. To date, CER facilitated ion diffusion has been reported for the synthesis of CSNPs with a variety of structures, such as amorphous/crystalline heterostructured CoV-Fe hydroxides[4], amorphous layered $Fe_3O_4$[5], and Au@$Ag_2S$ with amorphous $Ag_2S$ shell[6].

Amorphous nanomaterials can exhibit enhanced surface charge transfer compared to perfect crystals[7], which can result in improved performance for many optical and catalytic applications. For example, Guo's group has reported amorphous ZnO nanocages and amorphous $TiO_2$ nanosheets with remarkable surface-enhanced Raman scattering (SERS) due to their enhanced photo-induced charge transfer (PICT)[8]. Li's group reported that the high defect densities associated with amorphous components facilitate enhanced electrochemical performance for electronic oxygen reduction[9] and Zhang's group have reported that the coupled amorphous and crystalline components in Pd-P nanoparticles enhance their electrochemical performance[10]. Hence, the synthesis of nanomaterials with amorphous surface coatings is a promising area for engineering their improved function. Telluride NPs are valuable for a wide range of optoelectronic applications due to their suitable bandgap structure. However, tailoring the amorphous/crystalline structure of telluride NPs remains challenging compared to the sulfides and selenides, as telluride NPs possess weaker metal-Te bonds and easily deform when synthesized colloidally at high temperature[11, 12]. A facile strategy to engineer the surface of telluride NPs, including control of amorphous/crystalline structure, is therefore highly desirable[6a, 12a, 13].

In this work, we demonstrate surface tailoring achieved by aqueous CER. The synthesis of telluride NPs, having amorphous shells of variable thickness, was achieved by controlling the thermodynamics and kinetics of aqueous CER. Specifically, crystalline CdTe NPs (c-CdTe) and those covered with an amorphous shell (c@a-CdTe) were synthesized by aqueous CER, with the thickness of the amorphous shell tailored by control of the reaction temperature. A second CER step transformed the c@a-CdTe to crystalline CdTe@HgTe core-shell heterostructure NPs. SERS activities were measured to evaluate the optoelectronic properties of the NPs formed under different synthesis conditions. The as-prepared c@a-CdTe NPs exhibited superb SERS activity due to their high surface charge transfer enabled by the coupling of crystalline and amorphous components, as demonstrated by density functional theory (DFT) calculations.

## Results and discussion

The CER synthesis transforms $Ag_3AuTe_2$ NPs to crystalline CdTe NPs covered by amorphous shells (c@a-CdTe) with an adjustable shell thickness depending on the synthesis temperature (as illustrated in Fig. 1). High-resolution transmission electron microscopy (HRTEM), scanning transmission electron microscopy (STEM) annular dark field (STEM-ADF) imaging, STEM energy dispersive X-ray spectroscopy (STEM-EDX) and X-ray powder diffraction (XRD) were used to characterize the morphology, elemental distribution and crystallinity of the initial $Ag_3AuTe_2$ NPs. The $Ag_3AuTe_2$ NPs were synthesized by a hydrothermal method (see SI for full synthesis method). The aligned atomic lattice visible across the whole crystalline component of the particle in STEM-ADF images (Fig. S1A-C) indicates single crystal $Ag_3AuTe_2$ NP cores with high crystallinity covered by a thin amorphous shell. XPS spectra (Fig. S2) of the $Ag_3AuTe_2$ NPs indicates the existence of tellurium oxide, which can be expected to be in the amorphous shell. The XRD spectrum (Fig. S3B) matches the cubic phase $Ag_3AuTe_2$ (JCPDS No. 65-0444). Elemental maps measured by STEM-EDX (Fig. S1D-F) show a homogeneous distribution of Au, Ag and Te in the NP. Quantification of the STEM-EDX spectra summed over the whole particle reveals the elemental content of Te relative to Ag and Au is around 40 at%. This indicates an excess of Te compared to the stoichiometric ratio, also supporting the suggestion of a tellurium oxide shell, although this is below the detection limit of the STEM-EDX elemental maps.

Aqueous CER was applied on the $Ag_3AuTe_2$ NPs to form c@a-CdTe NPs. As previously reported, Ag-X (X=S, Se, Te) can be transformed to Cd-X under relatively mild conditions in a short time by phosphine-induced CER.[14] When the as-prepared $Ag_3AuTe_2$ NPs were reacted with cadmium tri-n-butyl phosphate (Cd-TBP) at 40°C, c@a-CdTe NPs were produced. The NPs had crystalline cores of diameter 46±3 nm and amorphous shells 7-8 nm thick (Fig. 2). Analysing the summed STEM-EDX spectra (Fig. S4A) for the c@a-CdTe NPs (acquired by averaging over a large viewing area) shows the presence of only 1 at% Ag and the complete absence of Au, which indicates that $Ag^+$ and $Au^+$ cations in the original NPs are almost entirely transferred to $Cd^{2+}$ by CER. The XRD pattern (Fig. S5A) shows only the crystalline cubic phase CdTe (PDF card number JCPDS No. 75-2086) present in the as produced NPs.

Detailed morphology, elemental distribution and crystallinity are confirmed by high-resolution STEM-ADF and STEM-EDX analysis. The as-synthesized CdTe NPs consist of a core-shell structure. The particle contains both Cd and Te, with the shell having a greater relative Te content (Fig. 2 A-D). The core shows high crystallinity with a measured lattice spacing matching the {111} planes of cubic phase CdTe (Fig. 2 E1). The outer edges of the NP in STEM-ADF images (Fig. 2 E2-4) show the diffuse intensity and an absence of atomic structure, which indicates that the NP shell is amorphous with a thickness of about 7-8 nm. Under the joint action of phosphorus ligands in the aqueous system, the CER can be driven by the thermodynamic temperature of 40°C. Nevertheless, the NPs are not fully crystallized due to the low temperature in the process of CER, and an amorphous layer is formed as an outer shell. Further surface XPS characterization suggests the presence of Te-O free bonds associated with the surface, which further supports the shell being amorphous (Fig. S6).

By adjusting the temperature of the aqueous CER, the amorphous shell thickness can be tailored as illustrated schematically in Fig. 1. When the reaction was performed at temperatures higher than 40°C, the thickness of the amorphous shell decreased, as demonstrated by HRTEM and STEM-EDX characterisation (Fig. 3 and Fig. S8). When reacted at 60°C, the as-prepared CdTe NP showed a highly crystalline core and an amorphous shell with a thickness around 4-5 nm (Fig. 3A-B). When the temperature was further increased to 80°C, the shell thickness further decreased (2-3 nm shell, as illustrated in Fig. 3C-D). When reacted at an even higher temperature (110°C), crystalline lattice could be detected by HRTEM imaging for the whole particle even near the surface (Fig. 3E-F), which suggests the amorphous shell is almost entirely removed. XPS spectra for the c@a-CdTe NPs synthesized at 110°C (Fig. S7) indicated a decrease of the Te-O signal compared to the XPS spectra of the c@a-CdTe NPs synthesized at 40°C (Fig. S6), which further supports the observed decrease of the amorphous shell thickness. Regardless of amorphous shell thickness, all as-prepared CdTe based NPs exhibited high monodispersity (Fig. S9), demonstrating that the aqueous CER method provides NPs with well controlled size distributions. The overall particle size is 53±2 nm for all synthesis temperatures, while the amorphous shell thickness reduces when synthesis temperature increase and the size of crystalline core correspondingly increases. It has previously been reported that fully crystalline CdTe quantum dots could be synthesized directly at 80 to 100°C within an aqueous system[15], indicating that the epitaxial growth of more CdTe on the c-CdTe cores is possible at higher temperatures, resulting in an enlargement of the size of the crystalline core. In addition, higher temperatures can be expected to increase the rate of ion diffusion through the amorphous surface shell, which will favour the supply of ions to the core.

To further demonstrate the flexibility of the CER strategy for tailoring the NP surface and core@shell morphology, a second spontaneous CER with $Hg^{2+}$ was applied to the as-prepared c@a-CdTe NP synthesized at 80°C. In recent years, many groups have reported the synthesis of $Hg_xCd_{1-x}Te$ and CdTe@HgTe quantum dots by CER between CdTe and $Hg^{2+}$ [16]. In our reaction system, when introducing excess $Hg^{2+}$ to the as-prepared c@a-CdTe (80°C), the $Hg^{2+}$ first reacted with the amorphous shell rather than the crystalline core (Fig. 4), which achieved a highly controlled c-CdTe@HgTe core-shell NP morphology as characterized by XRD, STEM-ADF and STEM-EDX analysis. The XRD spectrum matches well with cubic HgTe (PDF card number JCPDS No. 32-0665) as well as cubic CdTe (Fig. S5B), as both structures have similar diffraction patterns due to their similar crystal structure and lattice parameters. STEM-ADF results (Fig. 4A-B and S10A) demonstrate the high crystallinity of the as-produced NPs. The corresponding STEM-EDX results (Fig. 4C and S10B-D) show that the NP is fully crystalline although it has a CdTe core and a HgTe shell. The atomic lattice (Fig. 4B) extends continuously to the NP surface and matches the crystal structure expected for cubic phase CdTe (in agreement with XRD). No evidence is seen for extensive point defects or dislocations at the core-shell interface, which suggests the HgTe shell is smoothly transformed to the CdTe core. This epitaxy is possible due to the similar lattice parameter of CdTe and HgTe and relatively low bulk moduli of the two structures (445 and 423 kBar for CdTe and HgTe, respectively) [11c].

In order to evaluate the optical properties of the as-prepared NPs, their SERS activities when using 4-MBA as probe molecules were investigated. The UV-Vis tests of the colloidal of c@a-CdTe synthesized at 60°C, c-CdTe@HgTe, c-CdTe synthesized at 110°C and c@a-$Ag_3AuTe_2$ NPs and the corresponding 4-MBA modified NPs are shown in Fig. 5A-B. This UV-Vis spectral data demonstrated that 4-MBA modified c@a-CdTe NPs, c-CdTe@HgTe NPs, and c-CdTe NPs all exhibited an obvious broadband absorption with a slight red-shift compared to the control 4-MBA solution. The greatest increased absorption was observed for c@a-CdTe NPs indicating that these could exhibit the strongest photoinduced intermolecular charge transfer (PICT) resonance among the selected three samples, possibly due to their amorphous shell, in good agreement with the behaviour reported for ZnO and $TiO_2$ NPs[8]. The 4-MBA modified c@a-CdTe (60°C), c-CdTe (110°C), c-CdTe@HgTe NPs SERS spectra are shown in Fig. 5C. When irradiated by a 514.5 nm laser, all the three samples exhibited high SERS activity compared to the control 4-MBA spectra, and the intrinsic peak of 4-MBA exhibited no obvious shift. Among the characterized samples, c@a-CdTe (60°C) exhibited the highest SERS activity with a high calculated enhancement factor around $8.82×10^5$ (measured from 5 representative spectra acquired in different specimen areas, see Fig. S11 and Table S1). The c-CdTe and c-CdTe@HgTe NPs also exhibited SERS activity but with relatively lower activity. The higher EF value for c@a-CdTe (60°C) is hypothesized to be due to the high PICT enabled by the amorphous shell, as reported by Guo's group for other NP systems[7, 8, 17].

In order to investigate the potential effect of the coupling of crystalline and amorphous components on the EF value, the SERS measurements were performed for the other c@a-CdTe samples synthesized at different temperatures and found to have different amorphous shell thicknesses, as shown in Fig. 5D. The SERS results indicate that c@a-CdTe synthesized at 60°C (with a 5-6 nm thick amorphous shell) exhibited the highest SERS activity compared to c@a-CdTe synthesized at 40°C (with a 7-8 nm thick amorphous shell) and 80°C (with a 2-3 nm thick amorphous shell). It is hypothesized that the amorphous shell may provide enhanced PICT efficiency, while the crystalline core could provide a high density of photo-induced charge carries. The coupling of the amorphous shell and crystalline core could facilitate the enhanced SERS activity with optimal enhancement at the 5-6 nm shell thickness. All samples exhibited high uniformity and repeatability in the SERS data (Fig. 5C-D and Fig. S11 for 5 additional measurements). Compared to recently reported CSNPs SERS substrates (Table S2), the as-prepared c@a-CdTe (60°C) exhibited a relatively high enhancement factor up to $8.82 \times 10^5$ (Fig. S12), making them promising candidates for applications including spectroscopy and biochemical detection.

To study the interfacial charge transfer in the c-CdTe and a-CdTe systems, the distributions of charge differences in 4-MBA@c-CdTe and 4-MBA@a-CdTe were calculated by density functional theory (Fig. 5E-F). It was found that the a-CdTe surface facilitates a larger number of electrons (0.405 e) to transfer from the CdTe to the 4-MBA molecule, leading to a stronger PICT. The enhanced PICT could increase the 4-MBA electron density and thus change the molecular polarization. Hence, the isotropic polarizabilities of 4-MBA adsorbed onto the c- and a-CdTe surfaces were further calculated to evaluate the contribution of PICT to the molecular polarization. The results revealed that the a-CdTe surface could generate stronger molecular polarization (4-MBA@c-CdTe: 2494.03 $Bohr^3$/atom, 4-MBA@a-CdTe: 2585.16 $Bohr^3$/atom). Also, the calculated bandgaps of c-CdTe, a-CdTe bulk structures, and the PICT pathways in both 4-MBA@c- and a-CdTe systems indicated a favorable PICT in the 4-MBA@a-CdTe system as illustrated in Fig. S13-14.

## Summary

In summary, we have demonstrated a facile ACE strategy to synthesize telluride NPs with highly crystalline cores and thin amorphous shells of tunable thickness (c@a-CdTe NPs). By sequential CER we show that these materials can be transformed to core-shell telluride heterostructures (CdTe@HgTe core-shell NPs) with controlled morphology. All materials exhibited optical enhancement with the as-prepared c@a-CdTe synthesized at 60°C and having 5-6 nm amorphous shell exhibiting the highest SERS activity. The calculated enhancement factor was found to be around $8.82\times10^5$, indicating a high surface charge transfer ability, which is assigned to the presence of the amorphous shell with optimal thickness and its coupling to the crystalline core inducing a high density of photo-induced charge carriers.

## Supporting Information

Supporting Information is available from the Wiley Online Library or the authors.

## Acknowledgments

This work was supported by the National Natural Science Foundation of China (51631001, 52173232, 51902023, 51872030, 22005027 and 52072035), the Fundamental Research Funds for the Central Universities (2017CX01003) and China Postdoctoral Science Foundation (2020M670282). S.J.H and Y-C.W acknowledge funding from the Chinese Scholarship Council, the UK Engineering and Physical Sciences Research Council (EP/P009050/1 and EP/S021531/1) and European Research Council (Horizon 2020, grant agreement ERC-2016-STG-EvoluTEM-715502).

**Figures section:**

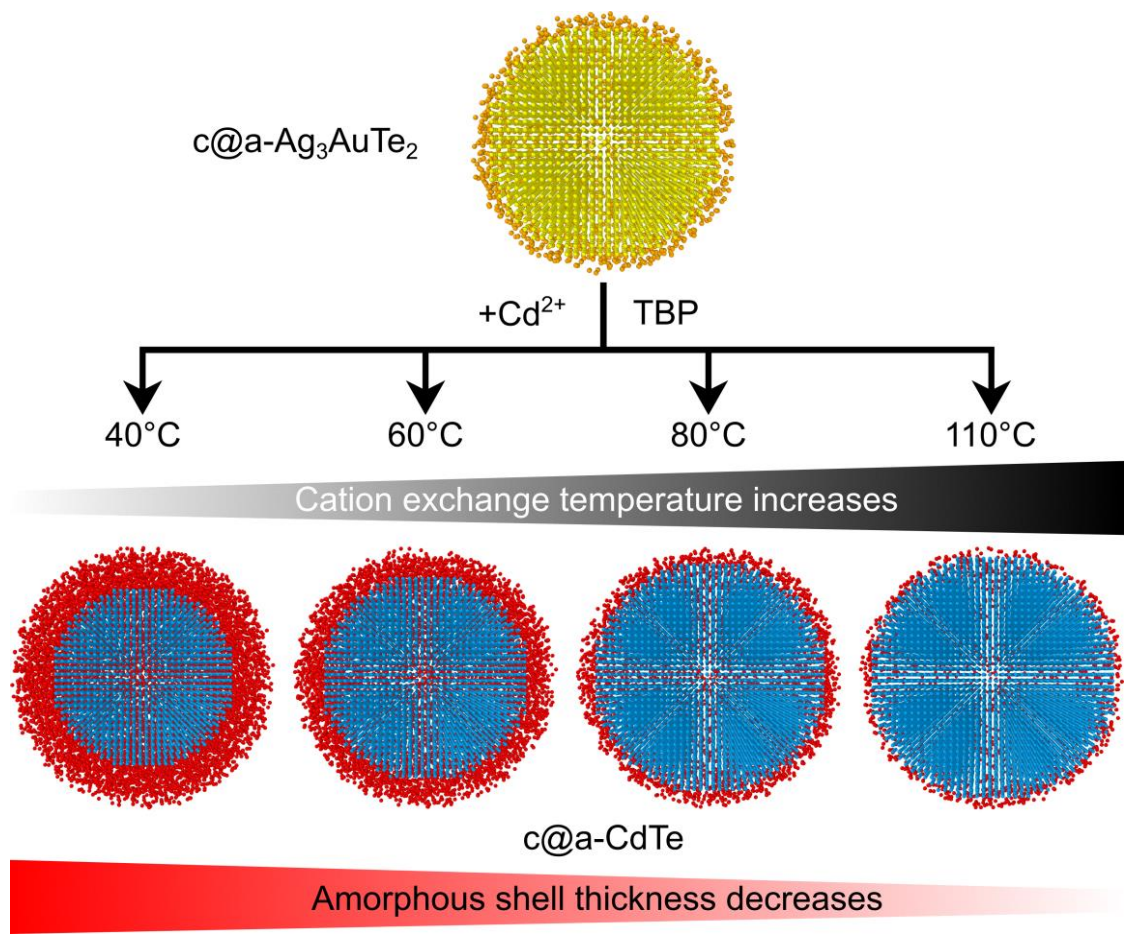

**Fig. 1. Schematic of the synthesis strategy.** Starting from crystalline $Ag_3AuTe_2$ NP, a cation exchange reaction is used to synthesize CdTe crystals with amorphous CdTe shells (c@a-CdTe) where the thickness of the amorphous shell decreases with increased reaction temperature from 40 to 110°C.

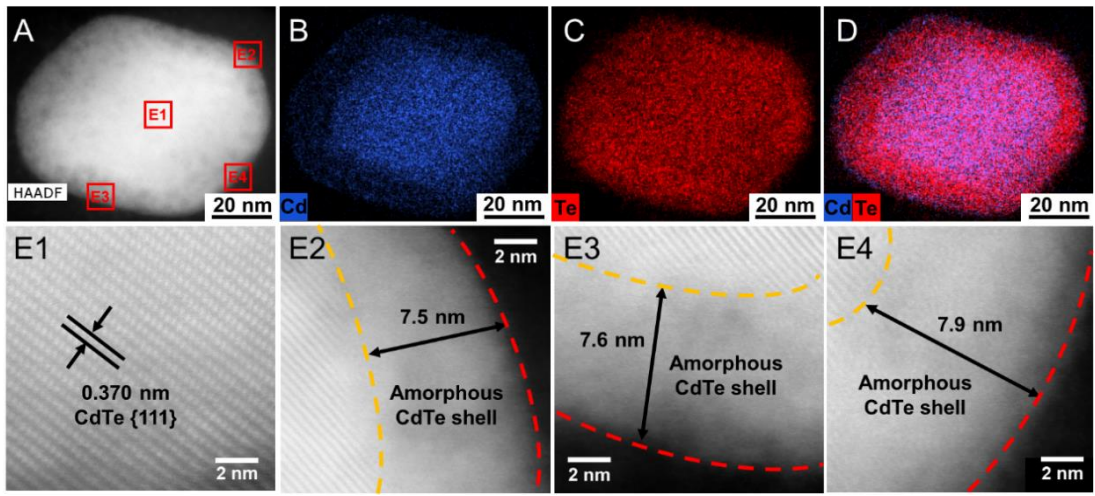

**Fig. 2. Characterization of c@a-CdTe NPs synthesized at 40°C**. HAADF-STEM image (A) and corresponding STEM-EDX elemental mapping for Cd and Te (B-D). Enlarged regions (E1-4) from (A) show that the nanoparticle possesses a highly crystallinity core and an amorphous shell with a uniform thickness of about 7-8 nm (E2-E4).

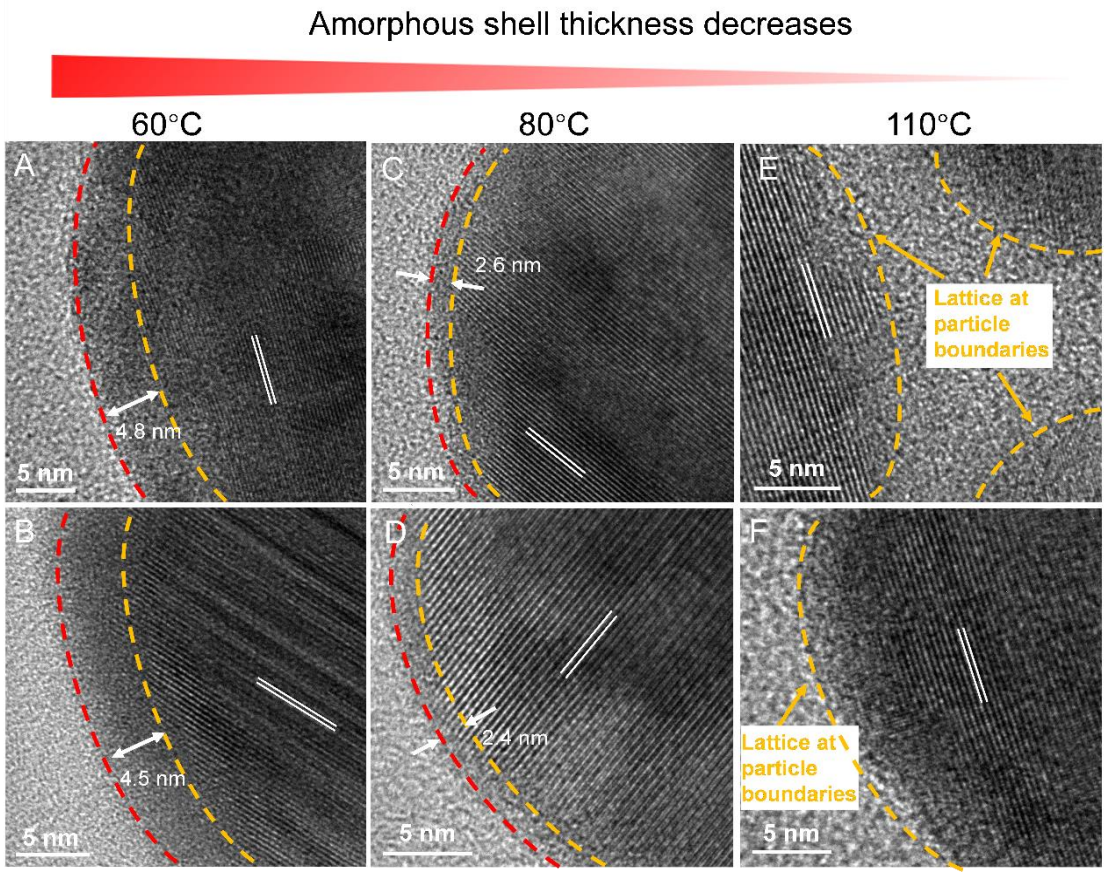

**Fig. 3. Characterization of c@a-CdTe NPs synthesized at 60°C, 80°C and 110°C.**

A-F) HRTEM images show the boundary of c@a-CdTe NPs synthesized at 60°C (A-B), 80°C (C-D) and 110°C (E-F). Yellow dashed lines mark the edge of the crystalline core, red dashed lines mark the edge of the amorphous shell. Lattice fringes are highlighted by parallel white lines and all matched the expected spacing of CdTe {111} lattice planes (0.370 ± 0.003 nm).

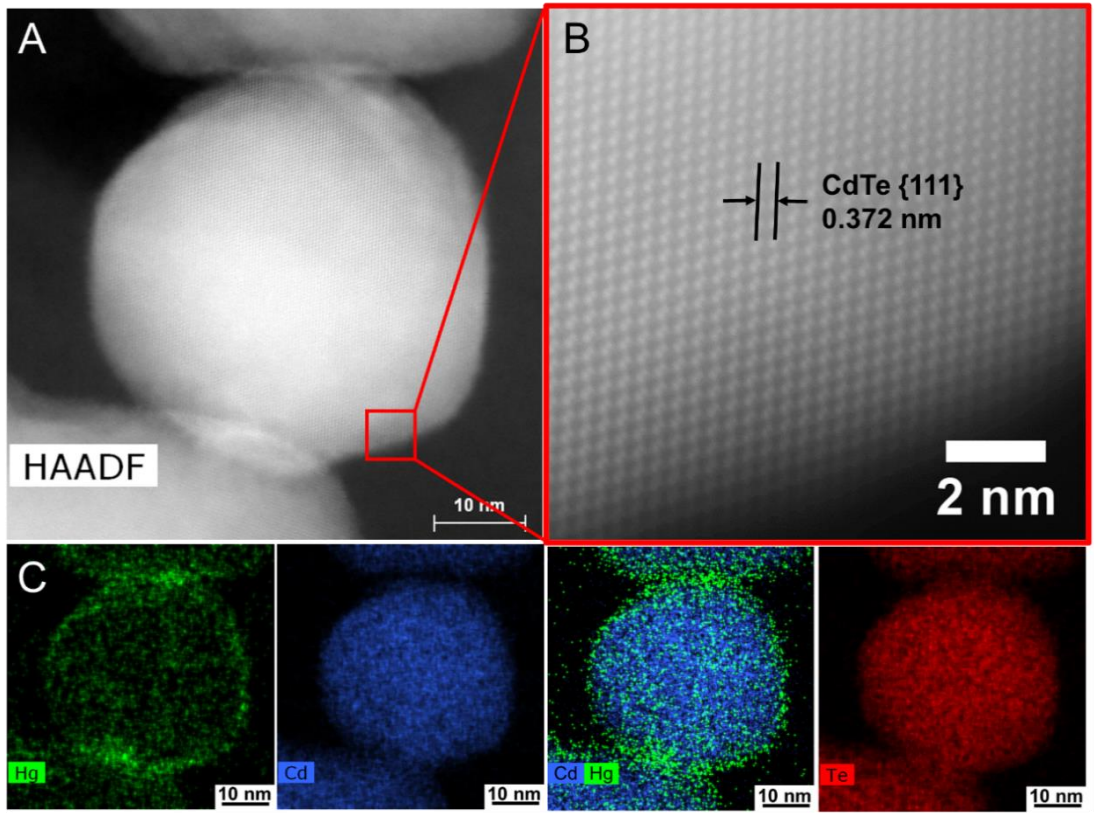

**Fig. 4. Characterization of c-CdTe@HgTe NP.** (A) STEM-ADF with (B) the enlarged region highlighted by the red square. (C) corresponding STEM-EDX maps of the region shown in (A).

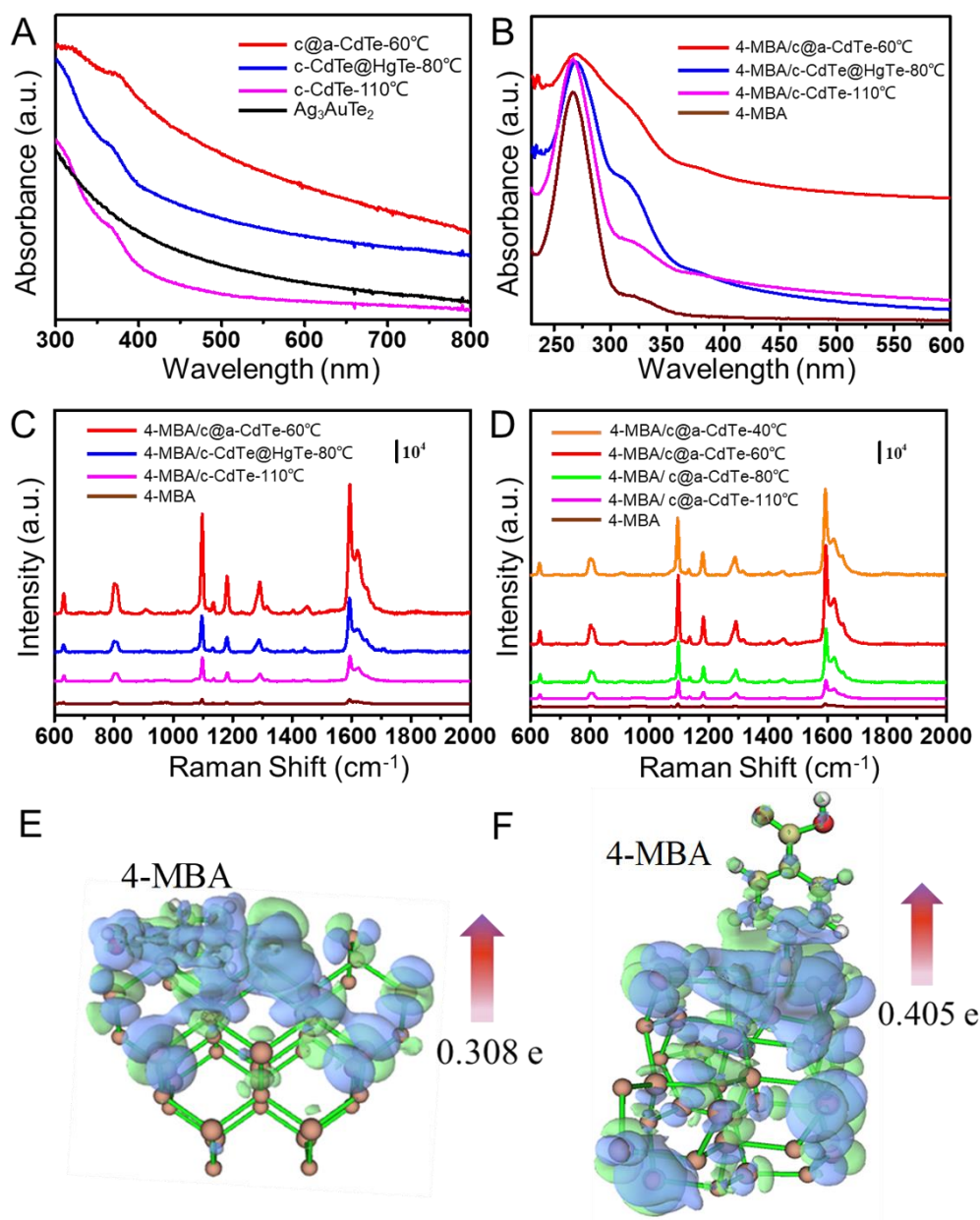

**Fig. 5. SERS responses and calculated charge differences distribution for the synthesized NPs.** (A) UV-Vis spectra of c@a-CdTe synthesized at 60°C, c-CdTe@HgTe, c-CdTe synthesized at 110°C and c@a-Ag$_3$AuTe$_2$ respectively. (B-C) UV-Vis spectra (B) and measured SERS spectra (C) of 4-MBA (0.01 M) molecules adsorbed on c@a-CdTe synthesized at 60°C, c-CdTe synthesized at 110°C and c-CdTe@HgTe. (D) Measured SERS spectra of mmv4-MBA modified CdTe NPs synthesized at different temperatures. (E-F) The difference in the charge distributions

for 4-MBA@c-CdTe and MBA@a-CdTe, respectively. The cyan (green) distribution illustrates electron depletion (accumulation). The electron transfer directions and values are also indicated.

# Supporting information:

## Telluride nanocrystals with adjustable amorphous shell thickness and core-shell structure modulation by aqueous cation-exchange

Xinyuan Li[a, b, +], Mengyao Su[a, +], Yi-Chi Wang[c, d +], Meng Xu[a, *], Minman Tong [e, *], Sarah J. Haigh[c], Jiatao Zhang[a, b, *]

a.  Beijing Key Laboratory of Construction-Tailorable Advanced Functional Materials and Green Applications, School of Materials, Beijing Institute of Technology, Beijing 100081, China. E-mail: zhangjt@bit.edu.cn; xumeng@bit.edu.cn

b.  School of Chemistry, Beijing Institute of Technology, Beijing 100081.

c.  Department of Materials, University of Manchester, Manchester M13 9PL, U.K.

d.  Beijing Institute of Nanoenergy and Nanosystems, Chinese Academy of Sciences, Beijing 101400, China

e.  School of chemistry and materials science, Jiangsu Normal University, Xuzhou 221116, China. Email: tongmm@jsnu.edu.cn

+   These authors contributed equally to this work.

**Chemicals:**

HAuCl$_4$·3H$_2$O, hydrazine hydrate (85%) and Te powder (99.999%) were purchased from Sinopharm Chemical Reagent Co., Ltd. Cd(NO$_3$)$_2$·4H$_2$O (99.99%) and hexadecyl trimethyl ammonium chloride (CTAC) were purchased from Aladdin Reagent. HgCl$_2$·4H$_2$O were purchased from Xiya chemical. Other chemicals were purchased from Beijing Chemical works. All reagents were used directly without further purification.

**Synthesis of crystalline@amorphous Ag$_3$AuTe$_2$ nanoparticles:**

The crystalline@amorphous Ag$_3$AuTe$_2$ (c@a-Ag$_3$AuTe$_2$) nanoparticles (NPs) were synthesized using a hydrothermal approach. First, Ag NPs were prepared following the previous reported seed-growth strategy with CTAC as surface ligands (2 ml ammonia was added to 5 mL 0.1M silver nitrate solution and stirred violently for 5 min. 10 mL with 0.2 M glucose solution was added and reacted in a reactor at 120°C for 9 h). Then 10 mL of the prepared Ag NPs colloidal solution was dissolved with 40 μL of 1 mM HAuCl$_4$·3H$_2$O solution followed by vigorous stirring for more than 30 minutes. Then 1.5 mL of aqueous Te precursor was prepared by heating a mixture of 30 mg Te powder and 15 mL of hydrazine hydrate in a 30 mL autoclave for more than 4 hours at 120°C, and this was introduced to the prepared Ag-Au solution. Then the whole mixture was vigorously stirred for 10 min followed by heating at 150°C for 4 hours in an autoclave. The mixture was then cooled to room temperature and the as-prepared Ag$_3$AuTe$_2$ NPs were collected by centrifugation at 7000 r/min for 10 min and re-dispersion in deionized water. The colloidal solution was then purified by a second cycle of centrifugation under the same conditions to wash out the residual Te precursor for subsequent cation exchange reactions.

**Synthesis of crystalline@amorphous CdTe nanoparticles:**

The core@shell crystalline@amorphous CdTe (c@a-CdTe) NPs were synthesized by phosphine induced aqueous cation exchange strategy. Typically, 10 mL of as-prepared c@a-Ag$_3$AuTe$_2$ colloidal was mixed with 1 mL of 0.2 M hexadecyl trimethyl ammonium bromide (CTAB) solution. Then, 1 mL of 30 mg/mL Cd(NO$_3$)$_2$ solution was added followed by vigorous stirring within 40°C water bath for more than 10 min. After that, 60 μL of tri-n-butylphosphine (TBP) was added under shaking and the whole colloidal was aged at 40°C in a water bath for 30 min. The as-prepared c@a-CdTe with 5 nm amorphous shell was then collected by centrifugation at 7000 r/min for 10 min and re-dispersed in deionized water. For the synthesis of c@a-CdTe NPs with different amorphous shell thicknesses, the synthesis procedures were similar except for changing the temperature of the water bath to 60°C or 80°C.

**Synthesis of crystalline-CdTe nanoparticles:**

The crystalline-CdTe (c-CdTe) NPs were synthesized by hydrothermal aqueous cation exchange reaction. Typically, 10 mL of as-prepared c@a-$Ag_3AuTe_2$ colloidal was mixed with 1 mL of 0.2 M CTAB solution. Then, 1 mL of 30 mg/mL $Cd(NO_3)_2$ solution was added followed by vigorous stirring within 80 °C water bath for more than 10 min. After that, 60 μL of TBP was added under shaking and the whole colloidal was transferred to a 30 mL autoclave followed by heating at 110°C for 2 hours. Then the autoclave was cooled down to room temperature and the as-prepared c-CdTe NPs were collected by centrifugation at 7000 r/min for 10 min and re-dispersed in deionized water.

**Synthesis of crystalline-CdTe@HgTe nanoparticles:**

The crystalline-CdTe@HgTe (c-CdTe@HgTe) NPs were synthesized by a spontaneous aqueous cation exchange reaction between CdTe NPs and $Hg^{2+}$. Typically, 10 mL of c@a-CdTe NPs with 2 nm amorphous shell were mixed with 1 mL of 0.2 M CTAC solution in a 30 °C water bath under stirring for more than 10 min. After that, 200 μL of $HgCl_2$ solution (5 mg/mL) was added under vigorously shaking. Then the whole colloidal was aged in the 30°C water bath for 30 min. The as-prepared c-CdTe@HgTe NPs were collected by centrifugation at 7000 r/min for 10 min.

**Structure characterization:**

NPs in deionized water were ultrasonicated for 10 minutes and then drop casted onto holey carbon support grids and dried at room temperature. TEM images were collected by using a HITACHI H-7650 transmission electron microscope operating with an accelerating voltage of 80 kV. Aberration corrected STEM-ADF and STEM-EDX spectrum images were collected using a ThermoFischer Titan ChemiSTEM operated at 200 kV, 90 pA beam current, 50 μs dwell time, total acquisition times of 5-30 minutes. STEM-EDX quantification was performed using a standard-less Cliff-Lorimer analysis in the Bruker Esprit and the Oxford instruments AZtec software without absorption correction (C and O were not quantified due to their presence in the carbon support film). High resolution TEM images were collected using JEOL JEM-2100F field emission transmission electron microscope operated at 200 kV. XRD patterns were collected by a Bruker D8 X-ray diffractometer (scan rate 6 °/min). UV–Vis spectra were measured on a Shimadzu UV-3600 spectrophotometer.

**SERS measurements:**

We chose 4-MBA as the probe molecule for SERS measurements. The as-prepared c@a-CdTe, c-CdTe and c-CdTe@HgTe NPs were collected by centrifugation at 7000 r/min for 10 min followed by re-dispersed in 500 μL of 0.01 M 4-MBA solution (the solvent was 50 vol% methanol/water solution). Then 200 μL of the colloidal solution were pipetted out and dropped on a clean Si wafer. The substrate was heated at 50°C to accelerate the vaporisation of solvents and then was subjected to SERS measurements. The laser wavelength was selected at 514.5 nm with the power fixed at 25 mW. The lens was 100× objective and the acquisition time was 10 s. For each SERS measurement, we obtained ten SERS spectra in different randomly chosen positions on the substrate and then took the average values of them for the calculation of enhancement factor.

**Theory calculation details:**

**Molecular dynamics (MD) simulation**

Amorphous CdTe was obtained by running MD simulation of crystalline CdTe under a high temperature of 1300 K based on the Forcite Module in Material Studio using universal force field [1]. The crystalline CdTe structure (c-CdTe) was obtained according to CdTe data (JCPDS Card No. 75-2086). A cutoff distance was set to 14.0 Å for the non-bonded interactions, while the long-range electrostatic interactions were handled using the Ewald summation technique.

**Bandgap calculation**

The bandgap was calculated by density functional theory (DFT) at the plane-wave level using CASTEP[2]. The exchange-correlation energy was treated in the generalized-gradient approximation (GGA) using the Perdew-Burke-Ernzerh (PBE) method [3]. As the experimental STEM data (Fig. 4) suggests a dominance of the {111} surface facets in the c-CdTe NPs (Fig. 4), the nano-model was constructed for c-CdTe and a-CdTe with 20 Å of added vacuum perpendicular to the (111) plane.

The calculations were carried out with the (3 × 3 × 1) Monkhorst-Pack k-points using a 330 eV cutoff energy. The (111) surface was modeled as a slab containing 48 and 50 atoms for c-CdTe and a-CdTe, respectively.

**The calculation of Static Interfacial Charge Transfer**

DFT calculations used the Gaussian 09 package to study the static interfacial charge transfer process. Initially, the 4-MBA molecule was put onto the 1st layer of the CdTe model by interacting the S atom in 4-MBA with the Cd atom in CdTe. The geometry optimizations were performed using B3LYP with the 6-31+G* basis set under standard convergence criteria [4]. The lanl2dz basis function was applied to the CdTe [5]. Multiwfn was used to analyze the charge transfer between the adsorbed MBA molecule and the CdTe. The simulation surface was modeled as a non-periodic slab with 43 atoms.

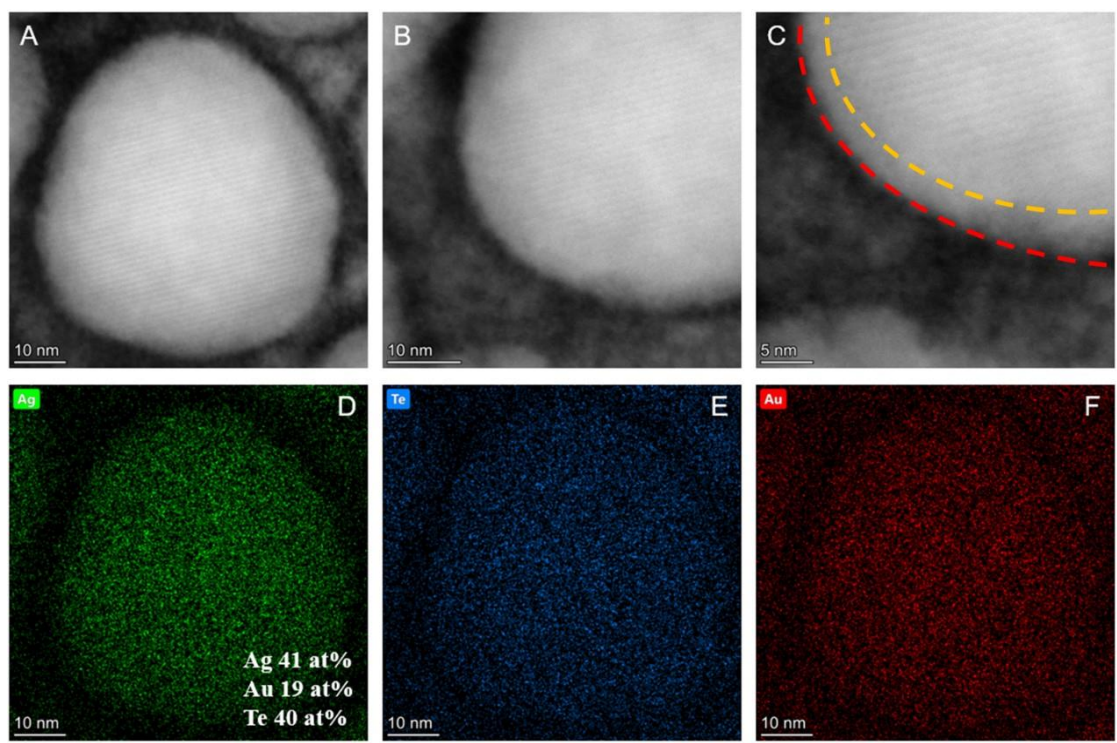

Fig. S1. STEM-ADF image (A-C) and STEM-EDX analysis results (D-F) for Ag, Te and Au respectively, of the as-prepared $Ag_3AuTe_2$ NP. The nanoparticles showed a single crystal structure and a thin amorphous shell (c@a-$Ag_3AuTe_2$), which is highlighted by the region between yellow and red dashed lines in F.

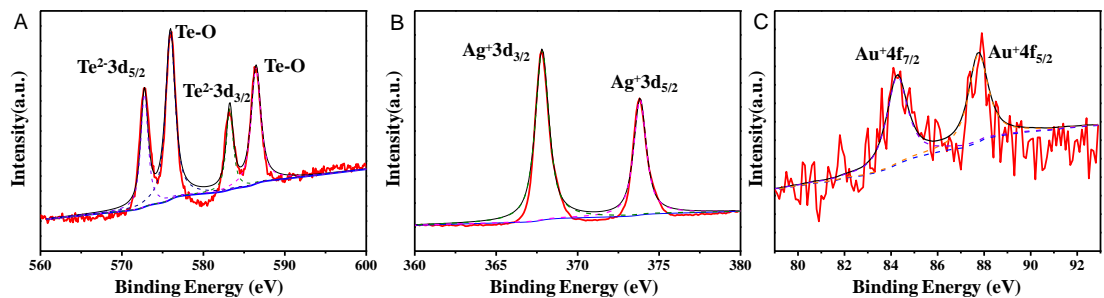

Fig. S2. XPS spectra for the c@a-Ag$_3$AuTe$_2$ NCs synthesized at 40°C.

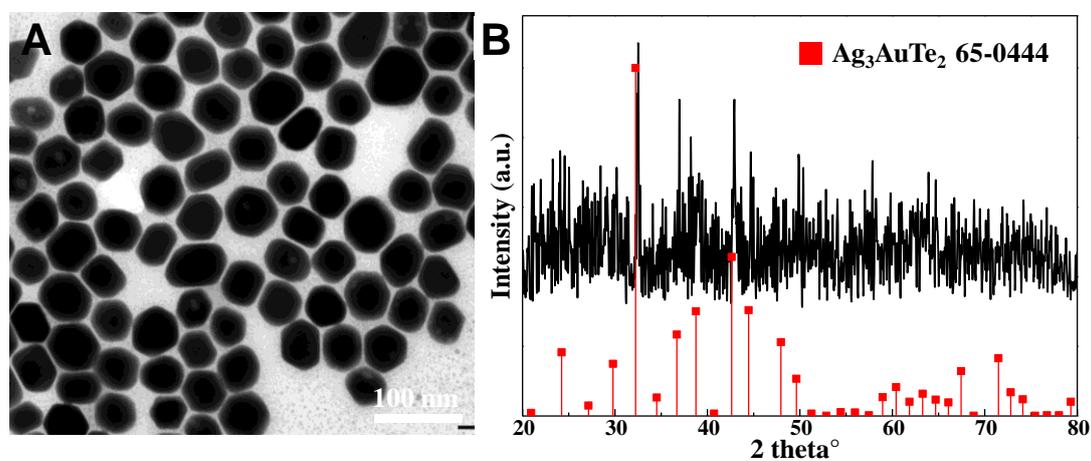

Fig. S3. TEM image (A) and XRD spectrum of the as-prepared c@a-Ag$_3$AuTe$_2$ NPs.

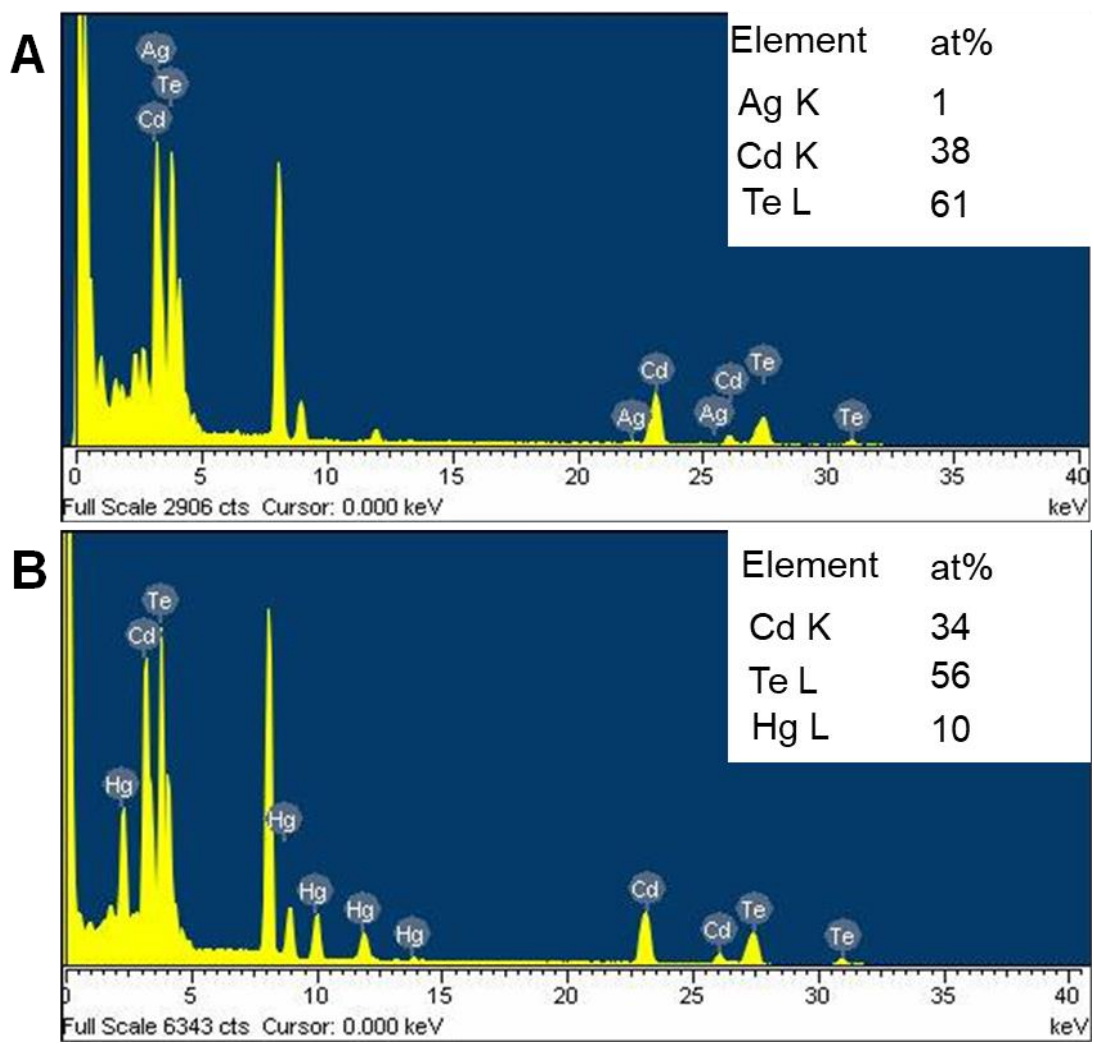

Fig. S4. Summed STEM-EDX spectra acquired over a large specimen area for the as-prepared c@a-CdTe NPs (A) and CdTe@HgTe NPs (B). Errors on Cliff-Lorimer quantification are ~10%. The peaks at 8.0 and 8.9 keV are Cu due to scattering from the TEM support grid.

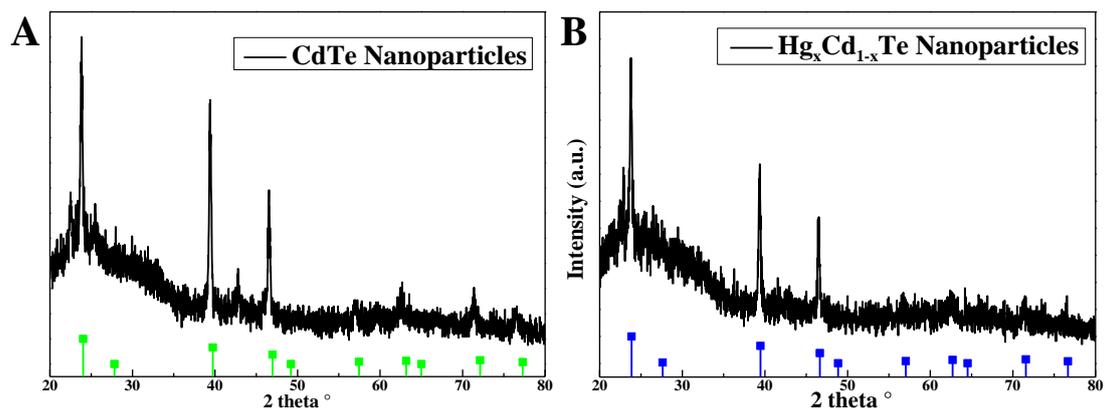

Fig. S5. XRD spectra of the as-prepared c@a-CdTe NPs (A) and CdTe@HgTe NPs (B). Green and blue reference spectra refer to cubic CdTe and cubic HgTe in (A) and (B) respectively (PDF card numbers JCPDS No. 75-2086 and JCPDS No. 32-0665)

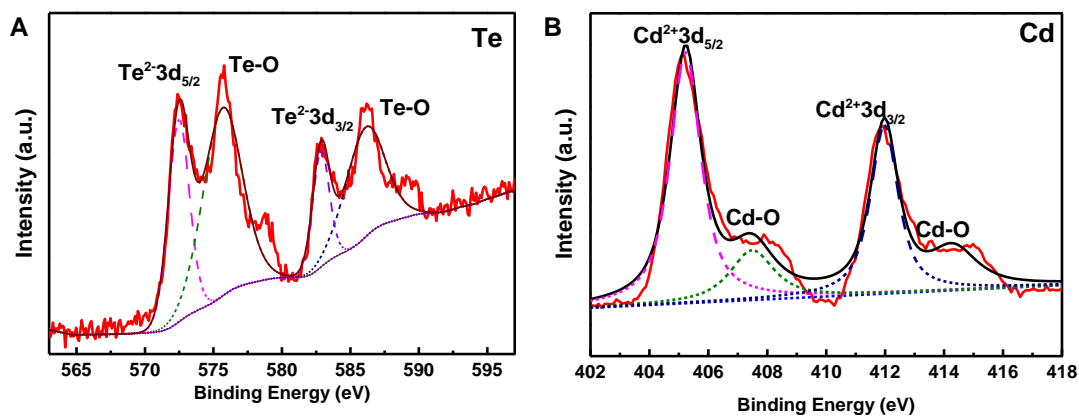

Fig. S6. XPS spectra from the c@a-CdTe NCs synthesized at 40°C.

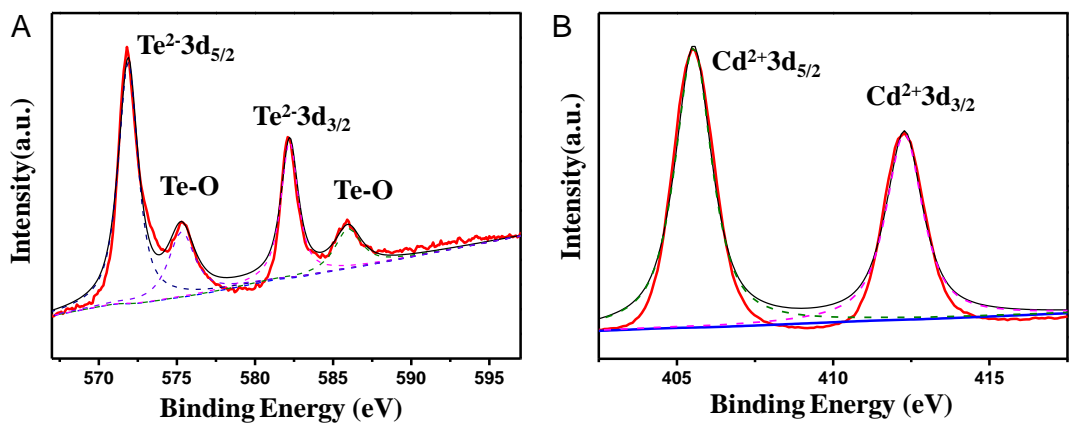

Fig. S7. XPS spectra from the c-CdTe NCs synthesized at 110°C.

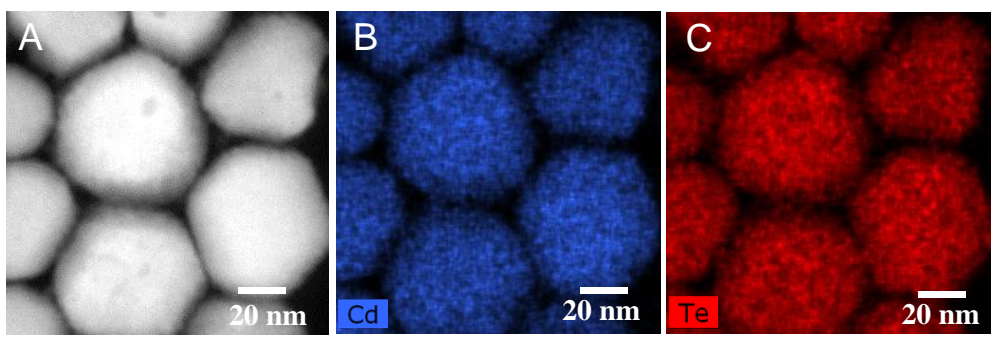

Fig. S8. STEM-EDX elemental analysis from c@a-CdTe synthesized at 80°C.

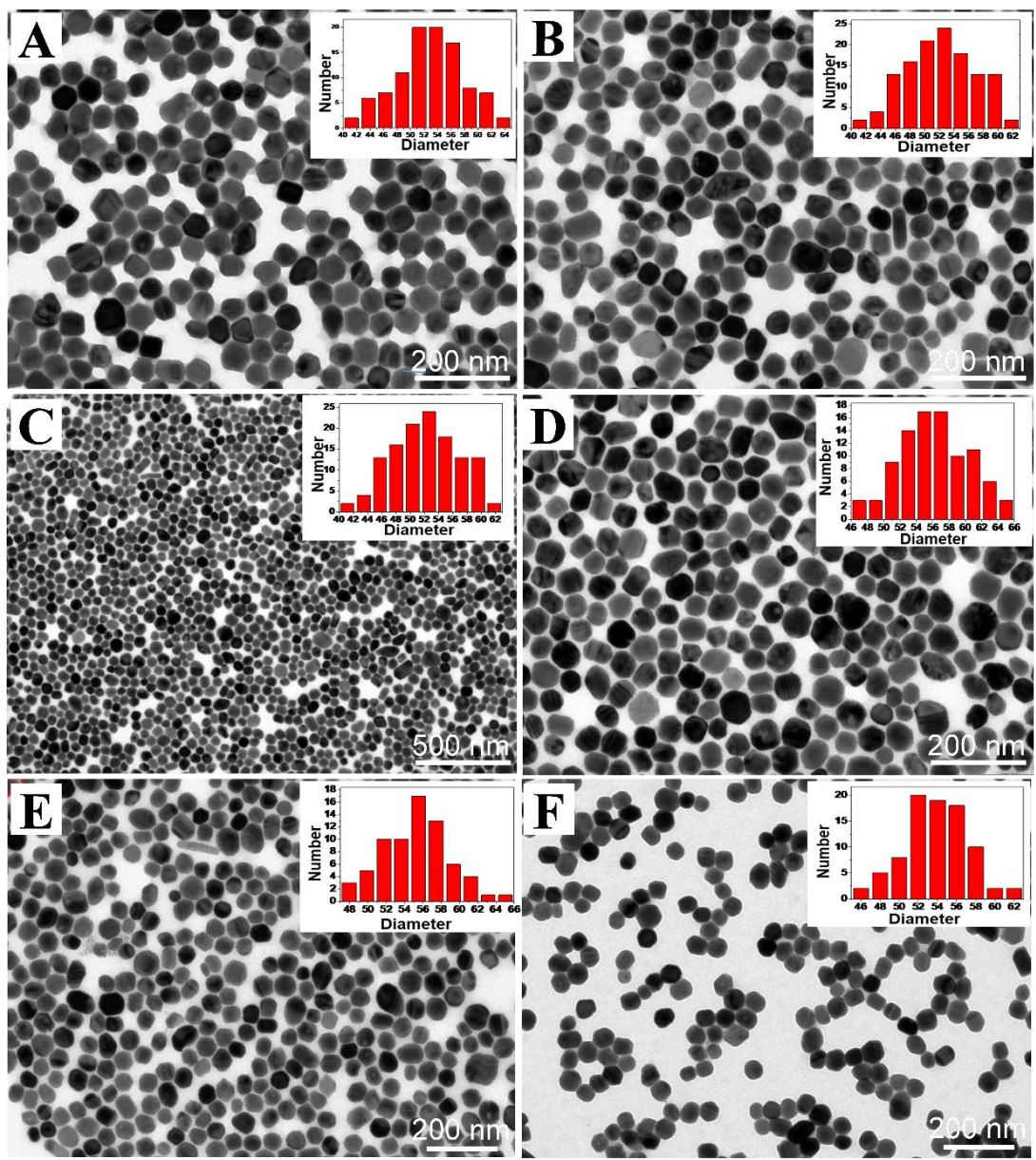

Fig. S9. TEM images and particle size map of as-prepared c@a-CdTe at 40°C (A), 60°C (B-C), 80°C (D) 110°C (E) and c-CdTe@HgTe NPs (F).

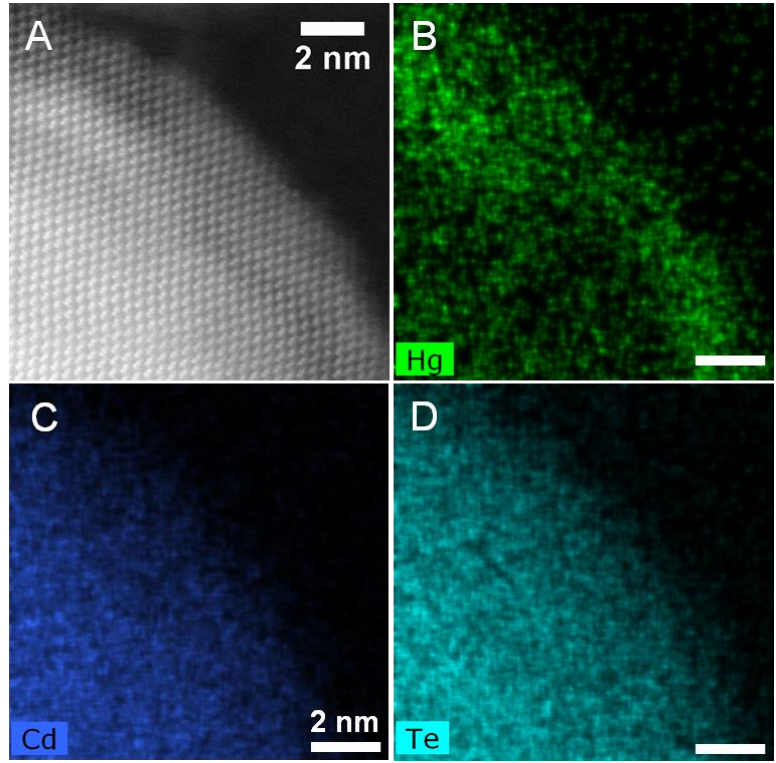

Fig. S10. STEM-ADF image (A) and STEM-EDX analysis results (B-D) of as-prepared CdTe@HgTe NP. Note the slight curvature of the lattice seen in (A) is due to mechanical instability during slow image acquisition.

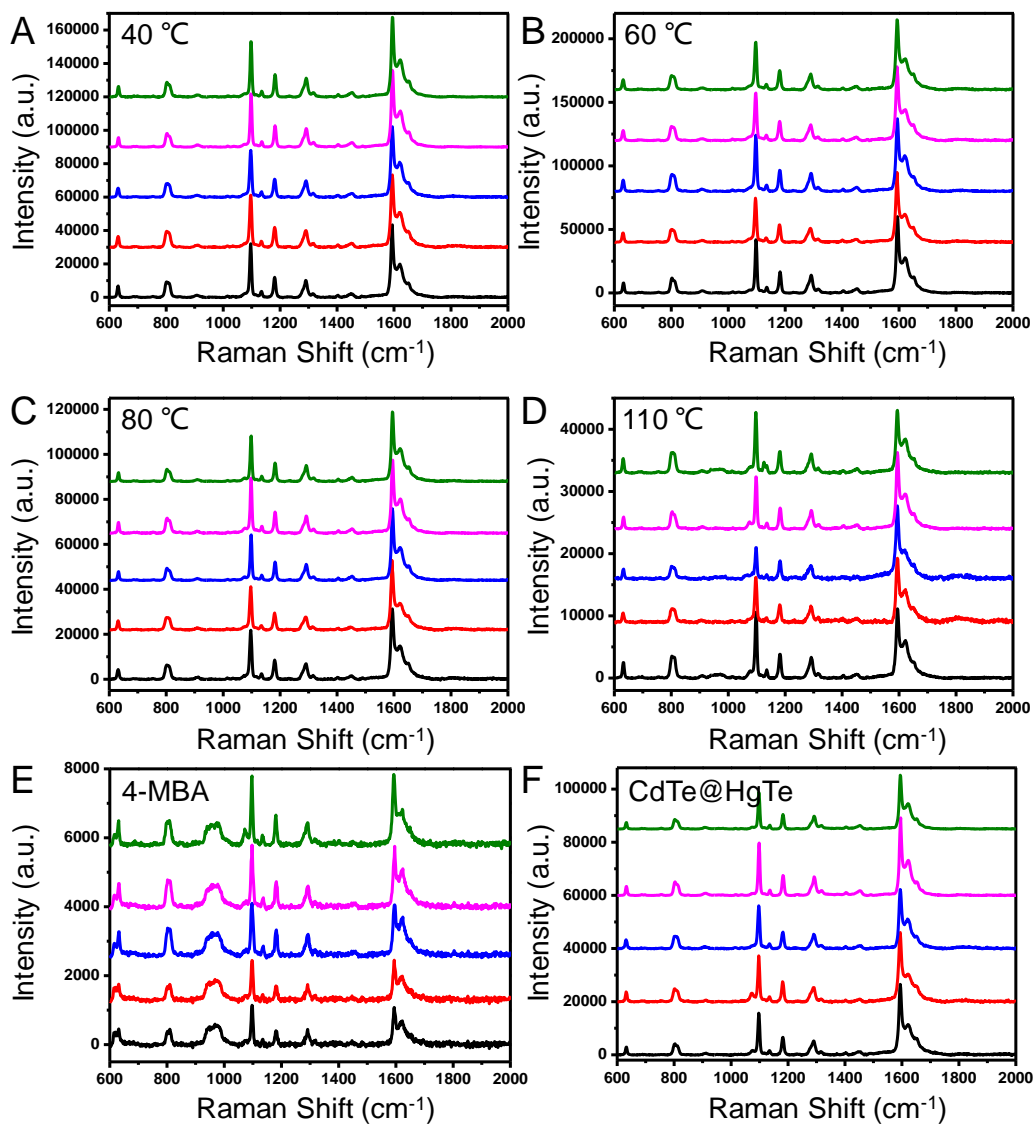

Fig. S11. Five representative SERS spectra (from different specimen areas) for 0.01M 4-MBA modified c@a-CdTe synthesized at 40°C (A), 60°C (B) and 80°C (C), c-CdTe synthesized at 110°C (D), c-CdTe@HgTe (E) and 0.01M 4-MBA without CdTe NCs (F) respectively.

# Calculation of SERS enhancement factor (EF) of c@a-CdTe NPs

The enhancement factor (EF) for c@a-CdTe NPs was calculated according to the following equation:

$$EF = (I_1/N_1) / (I_2/N_2) \quad (1)$$

where $I_1$ and $I_2$ denote the intensity of the surface (SERS) vibration peaks and Raman vibration peaks, respectively. $N_1$ and $N_2$ denote the number of 4-MBA molecules on the c@a-CdTe NPs substrate and in the probed volume of solution, respectively.

During the experiment, 200 μL of 4-MBA solution (0.01 M) was dropped onto the Si wafer (0.5×0.5 cm$^{-2}$). $N_2$ was thus estimated by:

$$N_2 = 200 \text{ μL} \times 0.01 \text{ mol/L} \times 6.02 \times 10^{23} \text{ mol}^{-1} \times S_1 / S_2 \quad (2)$$

where $S_1$ is the incident laser spot size and $S_2$ is the Si wafer size. The incident laser spot size $S_1$ is estimated by $S_1 = \pi r^2$, where $r = (1.22 \lambda/N_A)/2$. $\lambda$ is the incident wavelength 514.5 nm, and the numerical aperture of the objective lens $N_A = 0.5$, thereby, $S_1 = 1.24$ μm$^2$, $S_2 = 0.25$ cm$^2$. Thus $N_2$ was estimated to be $5.97 \times 10^{10}$.

$N_1$ is determined by the laser spot area illuminating the 4-MBA@ c@a-CdTe NPs substrate and the density of the 4-MBA molecule on the surface. Assuming the c@a-CdTe NPs can be approximated as an ideal flat surface, $N_1$ can be estimated as:

$$N_1 = \sigma \times 1.24 \text{ μm}^2 \times 6.02 \times 10^{23} \text{ mol}^{-1} \quad (3)$$

where $\sigma$ is the density of 4-MBA molecule adsorbed onto the c@a-CdTe NPs substrate, which is estimated to be ~ 0.5 nM cm$^{-2}$ (3). Thus $N_1$ is calculated to be $3.73 \times 10^6$.

As an example, experimental data in Figure S11 gives $I_1 = 60098$ and $I_2 = 1079$, for c@a-CdTe synthesized at 60 ℃ which by substitution of these values into Eq. (1), gives EF = $8.91 \times 10^5$. Other examples of calculated values for EF are shown in Table S1.

Table S1. Five representative SERS results and calculated EF for CdTe nanoparticles produced under different synthesize conditions.

| Materials | Synthesis strategy | EF | EF(average) standard deviation (SD) |
|---|---|---|---|
| Crystalline@amorphous-CdTe nanoparticles | 40°C | $7.06 \times 10^5$ | $6.89 \times 10^5$ SD=$0.35 \times 10^5$ |
| | | $7.26 \times 10^5$ | |
| | | $6.82 \times 10^5$ | |
| | | $7.07 \times 10^5$ | |
| | | $6.25 \times 10^5$ | |
| Crystalline@amorphous-CdTe nanoparticles | 60°C | $8.91 \times 10^5$ | $8.82 \times 10^5$ SD=$0.23 \times 10^5$ |
| | | $8.75 \times 10^5$ | |
| | | $9.10 \times 10^5$ | |
| | | $8.92 \times 10^5$ | |
| | | $8.42 \times 10^5$ | |
| Crystalline@amorphous-CdTe nanoparticles | 80°C | $4.82 \times 10^5$ | $4.55 \times 10^5$ SD=$0.29 \times 10^5$ |
| | | $4.72 \times 10^5$ | |
| | | $4.64 \times 10^5$ | |
| | | $4.58 \times 10^5$ | |
| | | $3.99 \times 10^5$ | |
| Crystalline-CdTe nanoparticles | 110°C | $1.65 \times 10^5$ | $1.64 \times 10^5$ SD=$0.12 \times 10^5$ |
| | | $1.52 \times 10^5$ | |
| | | $1.73 \times 10^5$ | |
| | | $1.82 \times 10^5$ | |
| | | $1.49 \times 10^5$ | |
| Crystalline-CdTe@HgTe nanoparticles | 80°C | $3.93 \times 10^5$ | $3.62 \times 10^5$ SD=$0.39 \times 10^5$ |
| | | $3.86 \times 10^5$ | |
| | | $3.30 \times 10^5$ | |
| | | $4.00 \times 10^5$ | |
| | | $3.03 \times 10^5$ | |

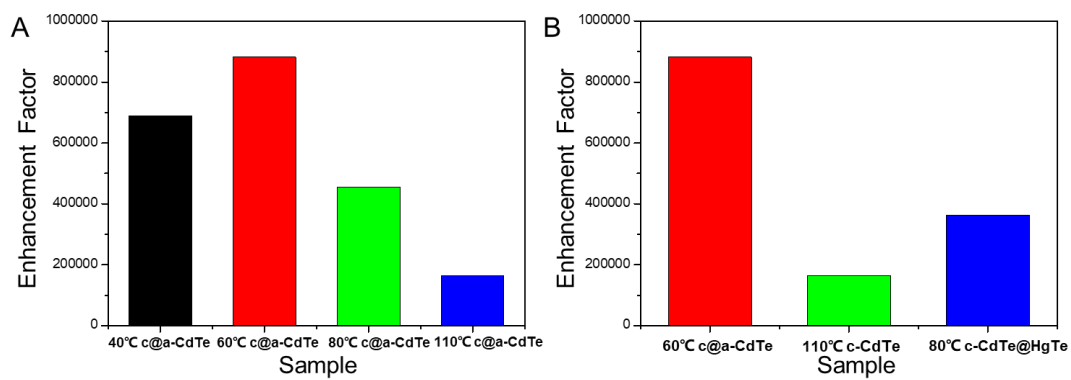

Fig. S12. The comparable SERS EF of as-prepared 4-MBA modified c@a-CdTe NPs (synthesized at different temperatures) and c-CdTe@HgTe. The measured SERS enhancement factors were obtained by taking the average of 5 measurements.

Table S2. Synthesis strategies, analytes, enhancement factors and applied laser wavelength of previously reported semiconductor nanomaterials with SERS characterization.

| Materials | Synthesis strategy | Analyte | EF | Laser(nm) | Reference |
|---|---|---|---|---|---|
| Amorphous $TiO_2$ nanosheets | High-temperature calcination | 4-MBA | $1.86 \times 10^6$ | 633 | Guo et al. *J. Am. Chem. Soc.* **2019**, *141*, 5856−5862. |
| Amorphous $Rh_3S_6$ microbowl | Liquid phase synthesis | R6G | $1 \times 10^5$ | 647 | Guo et al. iScience **2018**, *10*, 1–10. |
| Amorphous ZnO nanocage | High-temperature calcination | 4-MBA | $6.62 \times 10^5$ | 633 | Guo et al. Angew. Chem. Int. Ed. **2017**, *56*, 9851–9855. |
| Crystalline $MoS_{2-x}O_x$ nanosheet | High-temperature calcination | R6B | $1.6 \times 10^5$ | 532.8 | Zhang et al. Nat. Commun. **2017**, *8*, 1993. |
| $MoS_2$ nanoflower and nanosheet | High-temperature calcination | CA19-9 | $1.0 \times 10^5$ | 532 | Jiang et al. Biosens. Bioelectron. **2021**, *193*, 113481. |
| Yolk-shell Fe3O4@void@CeO2 | Liquid phase synthesis | MB | $1.1 \times 10^4$ | 488 | **Wang et al.** Appl. Surf. Sci., **2021**, *541*, 148422. |
| Crystalline $TiO_2$ inverse opal microarray | High-temperature calcination | MB | $2.0 \times 10^4$ | 532 | Zhang et al. J. Am. Chem. Soc. **2014**, *136*, 9886−9889. |
| Crystalline $Cu_2O$ nanosphere | Liquid phase synthesis | 4-MBA | $\sim 10^4$ | 488 | **Guo et al. Nanoscale, 2013**, *5*, 2784-2789. |
| **Crystalline@amorphous -CdTe nanoparticle** | **Aqueous cation exchange strategy** | 4-MBA | $8.82 \times 10^5$ | **514.5** | **This work.** |

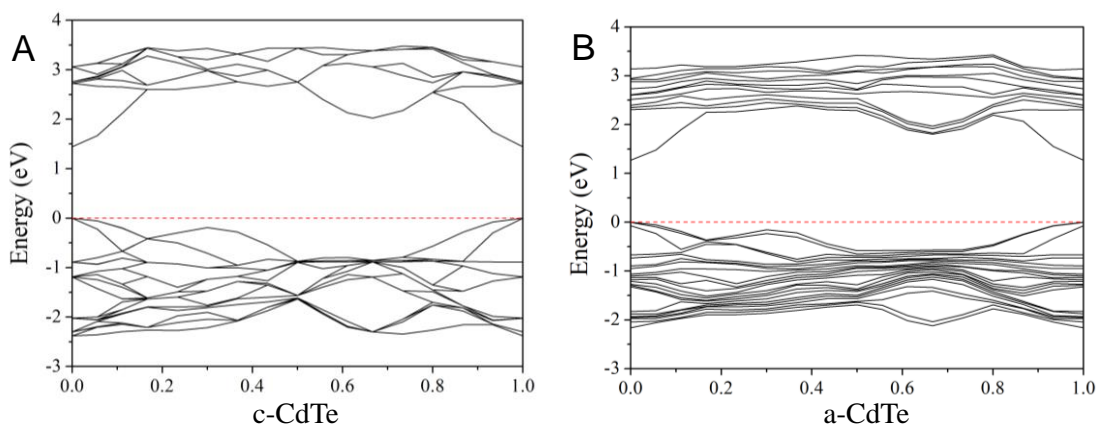

Fig. S13. Calculated band structures for (a) c-CdTe, (b) a-CdTe bulk structure.

For the bulk structures of c-CdTe and a-CdTe, the calculated band gaps are 1.445 eV and 1.268 eV, respectively (see Figure S13).

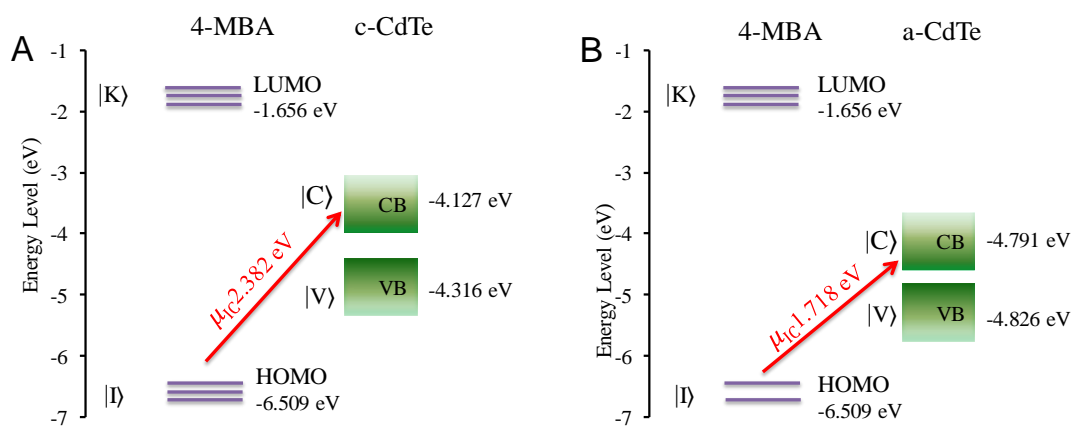

Fig. S14. Schematic energy-level diagrams of 4-MBA on the (a) c-CdTe and (b) a-CdTe surfaces.

As can be seen from Fig. S14, vibronic coupling is easier in the 4-MBA@a-CdTe system compared with the 4-MBA@c-CdTe system due to the closer valence band (VB) energy level value of a-CdTe to the HOMO (highest occupied molecular orbital) value of 4-MBA. $\mu_{IC}$ represents the energy of charge-transfer transition from the molecular ground states ($|I\rangle$) to the semiconductor conduction band (CB) states ($|C\rangle$) via the transition moment. $\mu_{IC}$ in 4-MBA@c-CdTe system (2.382 eV) is higher than that in 4-MBA@a-CdTe system (1.718 eV), revealing the favorable photoinduced charge transfer (PICT) in the 4-MBA@a-CdTe system.